# Analysis of potential locations of asteroidal moonlets[*]

Hexi Baoyin,[1]★ Xiaodong Liu,[1] and Laurène Beauvalet[2]

[1] *School of Aerospace, Tsinghua University, Beijing 100084, China*

[2] *IMCCE-Observatoire de Paris, 77, avenue Denfert-Rochereau, 75014 Paris, France*

## ABSTRACT

In this study, the potential locations of asteroidal small satellites (also called moonlets) with quasi-circular mutual orbit are analyzed. For the motion of the moonlets, only the solar gravity perturbation and the primary's 2nd degree-and-order gravity field are considered. By eliminating of short periodic terms, the dynamical behavior of the Hamiltonian for the moonlets is investigated. The observational data of some high size ratio binary asteroids show that the orbits of the moonlets lie close to the classical Laplace equilibria, which reach global minimum values of the Hamiltonian. It is found that tides or Yarkovsky effects alone cannot account for the reason why the orbits of asteroidal moonlets are not exactly at the classical Laplace equilibria. The analysis in this study is expected to provide useful information for the potential locations of asteroidal moonlets, and contribute to principles to relate predictions to observations.

---

# 1. Introduction

Binary minor planets are recent discoveries. The first confirmed binary asteroids 243 Ida-Dactyl were discovered in 1993 (Chapman et al. 1995; Belton et al. 1995, 1996). The investigations of binary minor planets have aroused great interest (Richardson & Walsh 2006). A comprehensive online database for binary asteroid systems is available on web page http://www.asu.cas.cz/~asteroid/binastdata.htm, the construction of which is described in Pravec & Harris (2007) and Pravec et al. (2012).

For the dynamics of binary asteroid systems, some work has been done in previous studies. The Generalized Tisserand Constant was used to elucidate orbital dynamical properties of distant moons of asteroids (Hamilton & Krivov 1997). In order to study the stability of the binary asteroids, the system was modeled based on the full two-body problem (Scheeres 2002a, b, 2004, 2006, 2007, 2009; Breiter et al. 2005; Fahnestock & Scheeres 2006). A two-dimensional dynamical model of the binary asteroids including primary's oblateness, solar perturbations and the BYORP (binary Yarkovsky-O'Keefe-Radzievskii-Paddack) effect enabled to obtain new results about orbital evolution (Ćuk & Nesvorný 2010). Numerical simulations were applied to investigate the stability of the binary asteroids 243 Ida (Petit et al. 1997), and the triple asteroids 87 Sylvia (Winter et al 2009; Frouard & Compère 2012). Both the stability regions around the triple asteroids 2001 SN263 (Araujo et al. 2012) and the collisionally born family about 87 Sylvia were also investigated using numerical models and integrations (Vokrouhlický et al. 2010). The Hill stability of binary minor planets was discussed using the total angular momentum and the total energy of the system (Donnison 2011). In our previous study, the Hill stability of triple minor planets was also examined (Liu et al. 2012). Scheeres et al. (2006) and Fahnestock & Scheeres (2008) studied dynamics of the near-Earth binary asteroids 1999 KW4. Fang et al. (2011) analyzed several processes that can excite the observed eccentricity and inclinations for near-Earth triple asteroids 2001 SN263 and 1994 CC. Further, Fang & Margot (2012) investigated the evolutionary mechanisms that can explain the origin of the spin with orbital parameters for near-Earth binaries and triples. Besides, there are plenty of papers on dynamics of a particle around an asteroid (Hamilton & Burns 1991, 1992; Chauvineau et al. 1993; Scheeres 1994; Scheeres et al. 1996, 2000; Rossi et al. 1999; Vasilkova 2005; Colombi et al. 2008; Yu and Baoyin 2012), which can also be applied to asteroidal moonlets.

The relevance of the dynamical behavior of asteroidal moonlets to the Laplace plane is studied in this study. Laplace (1805) introduced the concept of the Laplace plane of a planetary satellite[†]. For a satellite with circular orbit influenced by the planetary oblateness and the solar gravity perturbation, the Laplace plane is defined as the plane around which the instantaneous orbital plane of the satellite precesses. The Laplace plane possesses a constant inclination with respect to the planetary equatorial plane. The classical Laplace plane's axis is coplanar with and between the planet's spin axis and the planet's heliocentric orbit axis. In many works, dynamics of planetary satellites on the Laplace plane were studied. Allan & Cook (1964) found that for a circular orbit with given size, three mutually perpendicular directions in which the axis of the orbit remains stationary exist: two stable and one unstable. One of the stable directions corresponds to the classical Laplace plane. Ward (1981) showed circumplanetary disk's structure could affect the orientation of the local Laplacian plane. Stable rings are possible to exist in the circular orthogonal Laplace equilibria (Dobrovolskis 1980; Borderies 1989; Dobrovolskis et al. 1989a, b). Dobrovolskis (1993) studied the maps of Laplace planes for Uranus and Pluto, which are helpful for new satellites searches. Kudielka (1994) found that "balanced" Earth satellites orbits exist both in the classical Laplace plane and in the plane perpendicular to the classical Laplace plane. Tremaine et al. (2009) presented a comprehensive study of the Laplace equilibria including the effect of eccentricity. By truncating the gravitational potential up to the second order, Boué & Laskar (2006) presented the application of the Laplace plane to a three-body system consisting of a central star, an oblate planet, and a satellite orbiting the planet. Most of previous papers focused on the application of the Laplace plane to planetary satellites. Considering the two rigid bodies interactions, the concept of the Laplace plane was applied to binary asteroids to analyze the fully coupled rotational and translational dynamics (Fahnestock & Scheeres 2008; Boué & Laskar 2009). In Fahnestock & Scheeres (2008), the gravitational potential was expanded up to the second order, whereas in Boué & Laskar (2009), the gravitational potential was further expanded up to the fourth order.

---

[†] Sometimes, the term Laplace plane is used to refer to the invariable plane, the plane perpendicular to the angular momentum vector of the entire system.

Some high size ratio binary asteroids in the Solar System are found to possess quasi-circular mutual orbits, for example, 22 Kalliope, 45 Eugenia, 87 Sylvia, 107 Camilla, 121 Hermione, 216 Kleopatra. Recent studies were ever performed on the high size ratio binary asteroids. Four high size ratio main belt binary asteroids with quasi-circular mutual orbits were focused on in (Marchis et al. 2008). The evolution of the high size binary asteroids was studied using the MEGNO indicator and the truncated potential up to the second degree-and-order in (Compère et al. 2011). In this paper, only high size ratio binary asteroids with quasi-circular mutual orbits are considered, and simple model is used. People who are interested in more complicated models can refer to Boué & Laskar (2009) and Fahnestock & Scheeres (2008), which contributed significantly to the modeling of the binary asteroids. The analysis in this study is expected to provide a priori knowledge for the potential locations of asteroidal moonlets, and contribute to principles to relate predictions to observations.

## 2. The secular disturbing function due to the solar gravity perturbation and the primary's nonsphericity

In this study, we are concerned with high size ratio binary asteroids. The moonlet's effect on other bodies of the system is assumed negligible. This hypothesis comes from the fact that the moonlet's mass is expected to be too small to be detected, and as such, too small to have any major influence on the dynamics of the primary, which would end only in a very slight perturbation of the primary's heliocentric distance, and as such, a very small change in the solar gravity perturbation. In all the dynamical studies of such systems, as those of 45 Eugenia in Marchis et al. (2010) for example, the masses of the satellites have not been determined, but estimated from hypothesis on their density and size. Since the geometry of the moonlet would have an even smaller effect than the effect of its center of mass, we also do not take into account the shape of the moonlet. The orbit of the moonlet is under the influence of a variety of perturbations: the solar gravity perturbation, the solar radiation pressure, the gravitational harmonics of the primary, etc. The effects of these perturbations were analyzed in the previous research presented in the following. The solar radiation pressure affects significantly for very small particles, but slightly affects particles larger than a few centimeters (Hamilton & Burns 1992; Scheeres 1994). The gravitational harmonics dominate when close to the asteroid (Scheeres 1994). The solar gravity perturbation dominates when fairly far from the asteroid (Hamilton & Burns 1991; Scheeres 1994), and is important for the long-term evolution of the satellites (Yokoyama 1999). Thus, for the moonlet's motion, only the solar gravity perturbation and the nonspherical effect of the primary are considered. Further, we only consider the second degree-and-order gravitational harmonics for the nonspherical effect of the primary because of the large primary-moonlet separations with respect to the primary's radii. For simplicity, the mutual perturbations between moonlets are neglected if there are more than one moonlets in the system, the secular effects of which due to a secular resonance have been analyzed in (Winter et al. 2009).

In this paper, the primary's heliocentric orbital plane is taken as the reference plane. The perturbation due to the second degree-and-order gravity field from the primary is averaged with respect to both the primary's spin period and the moonlet's orbital period. The secular part of the disturbing function $R_p$ due to the primary's second degree-and-order gravity field in the primary's heliocentric orbital plane frame is obtained as (Kinoshita & Nakai 1991; Domingos et al. 2008; Tremaine et al. 2009)

$$R_p = n_p^2 J_2 R_e^2 \left[ 3(\cos i \cos \varepsilon + \sin i \sin \varepsilon \cos \Omega)^2 - 1 \right] / 4(1-e^2)^{3/2}, \tag{1}$$

where $n_p = \sqrt{\mu_p / a^3}$, $\mu_p$ is the primary's gravitational constant, $a$ is the moonlet's semimajor axis, $J_2$ is the oblateness coefficient, $R_e$ is the reference radius of the primary, $i$ is inclination, $\varepsilon$ is the inclination of the primary's equatorial plane with respect to the primary's orbital plane around the Sun, $\Omega$ is right ascension of the ascending node, and $e$ is eccentricity. Note that the gravity harmonic $J_{22}$ is eliminated by averaging over the primary's spin period.

The solar gravity perturbation is averaged with respect to both the primary's heliocentric orbital period and the moonlet's orbital period. The secular part of the disturbing function $R_\odot$ due to the solar gravity perturbation is (Kinoshita & Nakai 1991; Domingos et al. 2008; Tremaine et al. 2009)

$$R_\odot = \mu' n_\odot^2 a^2 \left[ (1+3e^2/2)(3\cos^2 i - 1)/8 + 15 e^2 \sin^2 i \cos 2\omega / 16 \right] / (1-e_\odot^2)^{3/2}, \tag{2}$$

where $\mu' = m_\odot / (m_p + m_\odot)$, $m_p$ is the mass of the primary, $m_\odot$ is the mass of the Sun, $n_\odot = \sqrt{\mu_\odot / a_\odot^3}$, $a_\odot$ is the primary's heliocentric orbital semimajor axis, $e_\odot$ is the primary's heliocentric orbital eccentricity, and $\omega$ is argument of pericentre.

The secular part of the disturbing function due to both perturbations is presented as

$$R = R_\odot + R_p. \tag{3}$$

## 3. The frozen solutions in the primary's heliocentric orbital plane

Based on the Lagrange's planetary equations (Chobotov 2002, p 201), the variation rates of $i$ and $\Omega$ can be easily derived as (Liu & Ma 2012)

$$\frac{di}{dt} = \frac{3n_p J_2 R_e^2}{2a^2}\sin\Omega \sin\varepsilon \left(\cos i \cos\varepsilon + \sin i \sin\varepsilon \cos\Omega\right), \tag{4}$$

$$\begin{aligned}\frac{d\Omega}{dt} = &-\frac{3\mu' n_\odot^2 \cos i}{4n_p \left(1-e_\odot^2\right)^{3/2}} - \frac{3n_p J_2 R_e^2}{4a^2 \sin i}\left(-\sin 2i \cos^2\varepsilon\right. \\ &\left. +\sin 2i \sin^2\varepsilon \cos^2\Omega + \cos 2i \sin 2\varepsilon \cos\Omega\right).\end{aligned} \tag{5}$$

For circular or quasi-circular moonlet's orbits, it is obvious that the Delaunay variables $L = \sqrt{\mu_p a}$ and $G = \sqrt{\mu_p a(1-e^2)}$ are constant. Thus, the averaged system has only one degree of freedom in $(H, \Omega)$, where $H = G\cos i$ (Murray & Dermott 1999, p 59). Since $G$ is constant, the orbital parameters $(i, \Omega)$ are used instead of the Delaunay variables $(H, \Omega)$. It is evident from Eq. (4) that there exist the frozen solutions when $\Omega = 0°$ (or 180°), which are the circular coplanar Laplace equilibria according to Tremaine et al. (2009). The values of these frozen $i$ for the Laplace equilibria can be solved numerically by setting the right-hand side of Eq. (5) equal to zero. Because either $\cos i$ or $\cos\Omega$ exist in the right-hand sides of both Eqs. (4) and (5), another two frozen solutions exist: $\Omega = \pm 90°$ and $i = 90°$, which are the circular orthogonal Laplace equilibria according to Tremaine et al. (2009). The linear stability of the Laplace equilibria including the oblateness and the solar gravity perturbation was examined using the vector description by Tremaine et al. (2009). In this paper, the stability of the Laplace equilibria to variations in $i$ and $\Omega$ is determined by analyzing the characteristic equation of the linearized model of Eqs. (4) and (5).

By defining a vector $X = (\delta i, \delta \Omega)^T$ as the variations, the variational equations of Eqs. (4) and (5) are written as

$$\dot{X} = \mathbf{A} \cdot X, \tag{6}$$

where

$$\mathbf{A} = \begin{bmatrix} \dfrac{\partial (\mathrm{d}i/\mathrm{d}t)}{\partial i} & \dfrac{\partial (\mathrm{d}i/\mathrm{d}t)}{\partial \Omega} \\ \dfrac{\partial (\mathrm{d}\Omega/\mathrm{d}t)}{\partial i} & \dfrac{\partial (\mathrm{d}\Omega/\mathrm{d}t)}{\partial \Omega} \end{bmatrix}.$$

The characteristic equation of Eq. (6) for the circular orthogonal Laplace equilibrium is calculated as

$$\lambda^2 = -\frac{9\mu' n_\odot^2 \sin^2 \varepsilon J_2 R_e^2}{8a^2 (1-e_\odot^2)^{3/2}}. \tag{7}$$

If $\lambda^2 > 0$, which means that one eigenvalue of $\mathbf{A}$ is a positive real number, so the Laplace equilibrium is unstable; if $\lambda^2 < 0$, both eigenvalues are pure imaginary, which means that the elements $i$ and $\Omega$ are both oscillatory, so the Laplace equilibrium is linearly stable. For some actual asteroidal moonlets, the examinations of stability to variations in $i$ and $\Omega$ will be also presented in Section 5.

## 4. Numerical verification

In this section, the averaged model is applied to 22 Kalliope's moonlet Linus for verification. The averaged results are compared to the direct numerical simulations of the full equations of motion including the unaveraged solar gravity perturbation and the unaveraged primary's second degree-and-order gravity field. The orbital parameters of the primary in the J2000 ecliptic coordinate system are available from the JPL Horizon service. The orbital elements of the moonlet in the J2000 Earth equatorial coordinate frame adopted in this paper are from Vachier et al. (2012). The derived spin vector solution of the primary in J2000 ecliptic coordinates is taken from Descamps et al. (2008). The primary's heliocentric orbital plane is adopted as the reference plane. The duration time of the orbital evolution is set to 10000 $T_s$.

The evolutions of the moonlet's orbital elements are presented in Fig. 1. It can be seen that the results of the averaged models show a satisfactory approximation to those of the unaveraged model for inclination $i$ and right ascension of the ascending node $\Omega$. Those two results are almost overlaid with each other. The mean $\Omega$ is about 0º and the mean $i$ is about 93.7º, which meets the frozen condition discussed in Section 3.

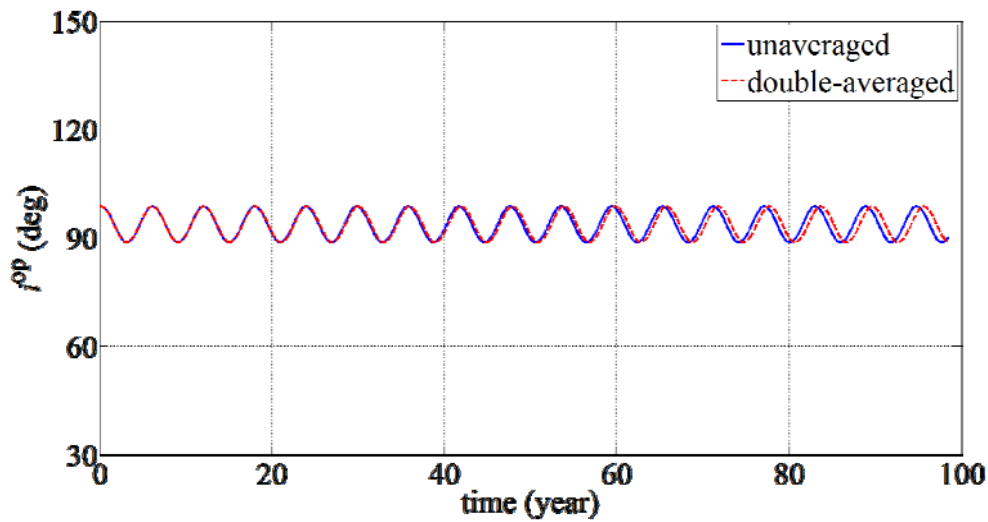

(**a**) Evolution of $i$ over 10000 $T_s$.

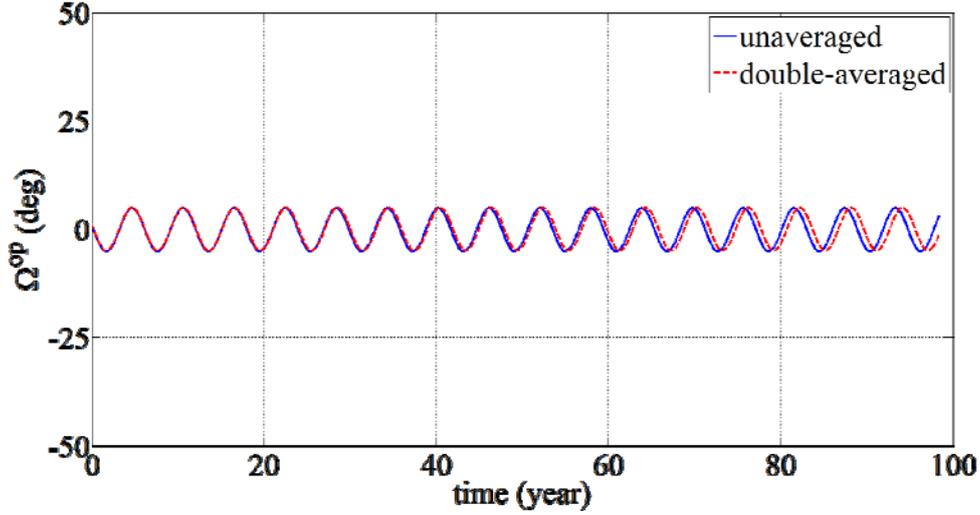

(**b**) Evolution of $\Omega$ over 10000 $T_s$.

**Fig. 1.** Evolutions of Linus's orbital elements over 10000 $T_s$. Solid line in blue correspond to results of the direct numerical simulations of the full (unaveraged) equations of motion; dashed lines in red correspond to results of the averaged model.

### 5. Analysis of locations of asteroidal moonlets

After averaging, the Hamiltonian can be presented as follows

$$\mathcal{H} = -\mu_p/2a - R. \tag{8}$$

It is obvious that the averaged Hamiltonian is time-independent for asteroidal moonlets with quasi-circular orbits, so the averaged Hamiltonian is an integral constant and represents the energy of the averaged system. Define the Hessian matrix $\mathbf{H}_s$,

$$\mathbf{H}_s = \begin{bmatrix} \dfrac{\partial^2 \mathcal{H}}{\partial i^2} & \dfrac{\partial^2 \mathcal{H}}{\partial i \partial \Omega} \\ \dfrac{\partial^2 \mathcal{H}}{\partial \Omega \partial i} & \dfrac{\partial^2 \mathcal{H}}{\partial \Omega^2} \end{bmatrix}. \tag{9}$$

If $\mathbf{H}_s$ is positive definite at the Laplace equilibrium, then the Hamiltonian $\mathcal{H}$ attains a local minimum at this equilibrium. If $\mathbf{H}_s$ is negative definite at the Laplace equilibrium, then $\mathcal{H}$ attains a local maximum at this equilibrium.

Several asteroidal moonlets 22 Kalliope's moonlet Linus, 121 Hermione's moonlet S/2001 (121) 1, and 45 Eugenia's moonlets Petit-Prince and Petite-Princesse are taken as examples to analyze the behaviors of the Hamiltonian in the parameter plane of $i$ and $\Omega$. The eccentricity for the moonlet' orbit is kept equal to zero, and the semimajor axis for the moonlet' orbit is kept as its actual value. The orbital parameters of Linus, S/2001 (121) 1, Petit-Prince, and Petite-Princesse are taken from Vachier et al. (2012), Descamps et al. (2009), Beauvalet et al. (2011), and Beauvalet et al. (2011), respectively. These moonlets' orbits are all almost-circular. The spin vector solutions of the primaries 22 Kalliope, 121 Hermione, and 45 Eugenia are taken from Descamps et al. (2008), Descamps et al. (2009), and Beauvalet et al. (2011), respectively. For the primary 22 Kalliope, $J_2 = 0.19$ (Descamps et al. 2008); for 121 Hermione, $J_2 = 0.28$ (Descamps et al. 2009); and for 45 Eugenia, $J_2 = 0.060$ (Marchis et al. 2010). It can be seen that the primary's $J_2$ is much larger than Earth's $J_2 = 1.08263 \times 10^{-3}$ (Lemoine et al. 1998) and Martian $J_2 = 1.95545 \times 10^{-3}$ (Lemoine et al. 2001). Simulations of 22 Kalliope's moonlet Linus, 121 Hermione's moonlet S/2001 (121) 1, and 45 Eugenia's moonlets Petit-Prince and Petite-Princesse are shown in Figs. 2-5, respectively.

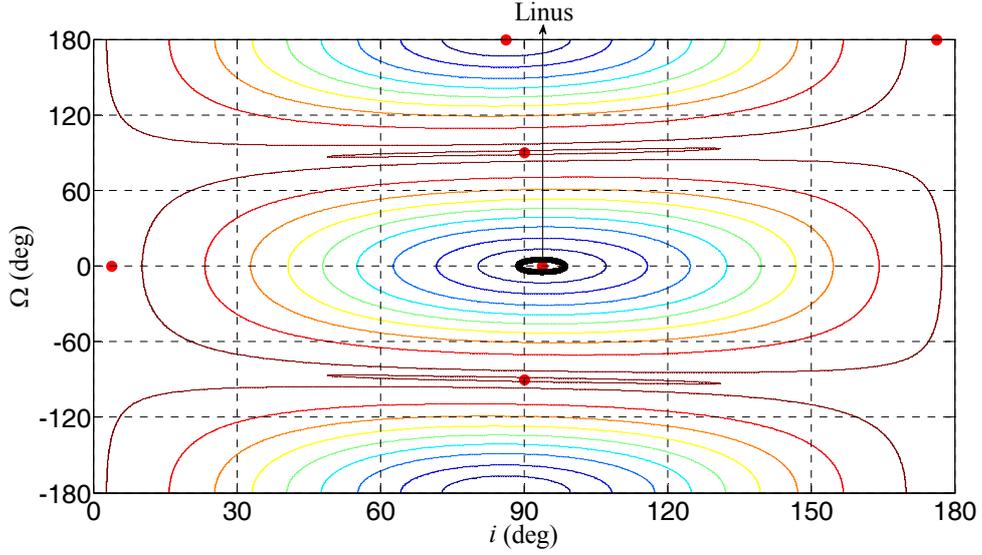

**Fig. 2.** Contours of the averaged Hamiltonian $\mathcal{H}$ in the parameter plane of $i$ and $\Omega$ for 22 Kalliope's moonlet Linus. The thicker line in black corresponds to Petite-Princesse's orbit of 10000 $T_s$. The red dots correspond to the Laplace equilibria.

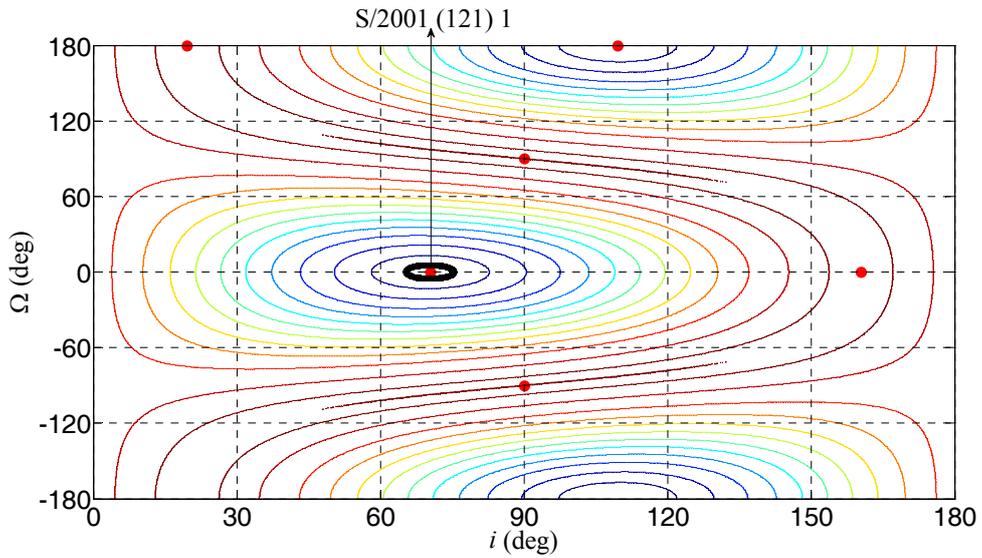

**Fig. 3.** Contours of the averaged Hamiltonian $\mathcal{H}$ in the parameter plane of $i$ and $\Omega$ for 121 Hermione's moonlet S/2001 (121) 1. The thicker line in black corresponds to S/2001 (121) 1's orbit of 10000 $T_s$. The red dots correspond to the Laplace equilibria.

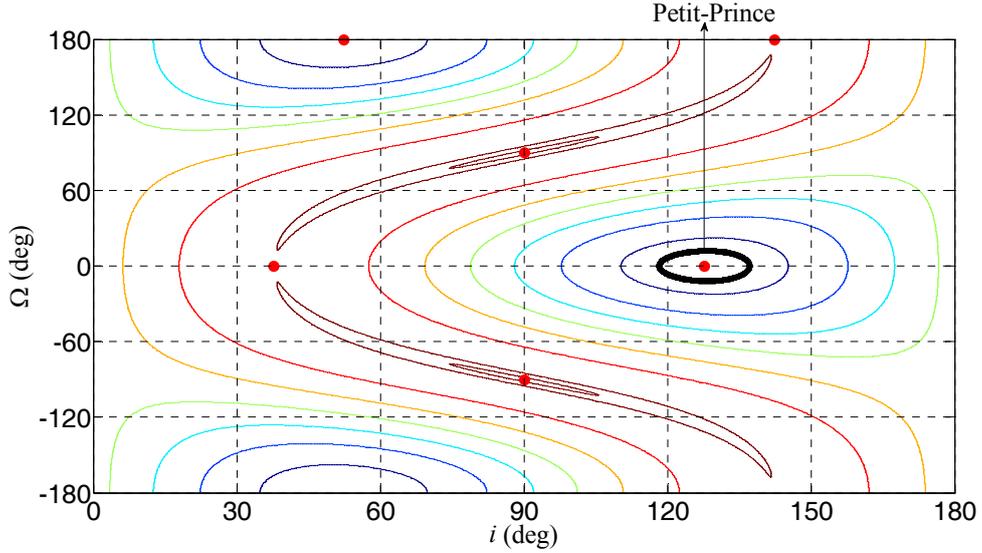

**Fig. 4.** Contours of the averaged Hamiltonian $\mathcal{H}$ in the parameter plane of $i$ and $\Omega$ for 45 Eugenia's moonlet Petit-Prince. The thicker line in black corresponds to Petit-Prince's orbit of 10000 $T_s$. The red dots correspond to the Laplace equilibria.

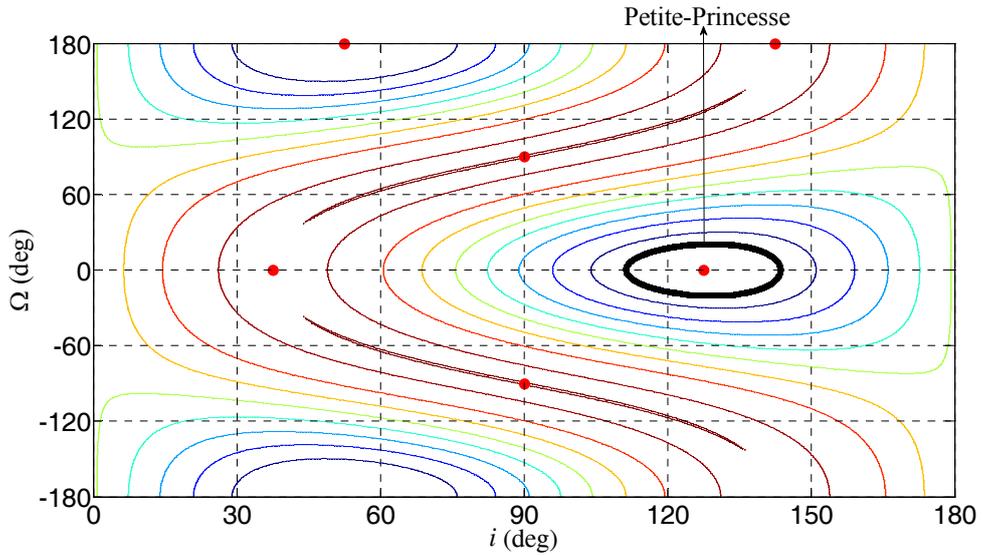

**Fig. 5.** Contours of the averaged Hamiltonian $\mathcal{H}$ in the parameter plane of $i$ and $\Omega$ for 45 Eugenia's moonlet Petite-Princesse. The thicker line in black corresponds to Petite-Princesse's orbit of 10000 $T_s$. The red dots correspond to the Laplace equilibria.

It is evident in Fig. 2 that six Laplace equilibria are found in total in the range of $i \in [0, 180º]$ and $\Omega \in [-180º, 180º]$ for 22 Kalliope's moonlet Linus. The values of the frozen $i$ and $\Omega$ at these Laplace equilibria can be obtained by solving equilibrium solutions of Eqs. (4) and (5), which are shown as follows

$(\Omega = 90°, i = 90°); (\Omega = -90°, i = 90°); (\Omega = 0°, i = 93.74°);$
$(\Omega = 180°, i = 86.26°); (\Omega = 0°, i = 3.74°); (\Omega = 180°, i = 176.26°).$

According to Tremaine et al. (2009), the equilibria when $i = 90º$ are the circular orthogonal Laplace equilibria, and the other four equilibria are the circular coplanar Laplace equilibria. For the first four Laplace equilibria, the eigenvalues of Eq. (6) are all pure imaginary according to Eq. (7), so the elements $i$ and $\Omega$ are all oscillatory. Thus, these four Laplace equilibria are all linearly stable to variations in $i$ and $\Omega$. For other two Laplace equilibria, one of the eigenvalues of Eq. (6) is a positive real number, so they are unstable to variations in $i$ and $\Omega$. Based on Eq. (9), the extremum properties of the linearly stable Laplace equilibria are examined. For the circular coplanar Laplace equilibria, the Hessian matrix $\mathbf{H}_s$ is negative definite according to Eq. (9), so the Hamiltonian $\mathcal{H}$ attains a local maximum at these equilibria. For the classical Laplace equilibrium $(\Omega = 0°, i = 93.74°)$ and the other circular coplanar linearly stable Laplace equilibrium $(\Omega = 180°, i = 86.26°)$, $\mathbf{H}_s$ is positive definite, so the Hamiltonian $\mathcal{H}$ attains a local minimum.

Seen from Figs. 3-5, there are also six Laplace equilibria for 121 Hermione's moonlet S/2001 (121) 1, and 45 Eugenia's moonlets Petit-Prince and Petite-Princesse: two linearly stable equilibria with local minimum values of the Hamiltonian, two linearly stable equilibria with local maximum values of the Hamiltonian, and two unstable equilibria. It is noted that from Figs. 2-5 that the orbits of these actual asteroidal moonlets all lie close to the classical Laplace equilibria that reach global minimum values of the Hamiltonian $\mathcal{H}$, which means that the normal of the averaged moonlet's orbital plane, the primary's spin axis, and the normal of the primary's heliocentric orbital plane are approximately coplanar. The reason why the orbits of asteroidal moonlets are not exactly at the classical Laplace equilibria might be due to the effect of the other perturbations. Yet, our knowledge of the dissipative forces in these kinds of systems suggests that tides or Yarkovsky effects alone cannot account for this. If we consider the tidal effects between 45 Eugenia and Petit-Prince, the primary's tidal Love number $k_p$ of 45 Eugenia is expected to be given by (Goldreich & Sari 2009),

$$k_p \approx 10^{-5} \frac{R_e}{1km}.  \qquad (10)$$

If we suppose that Petit-Prince is in a spin-orbit resonance with respect to 45 Eugenia, we can then have an estimation of its semi-major axis changing rate (Goldreich & Sari 2009),

$$\frac{1}{a}\frac{da}{dt} = 3\frac{k_p}{Q_p}\frac{m_m}{m_p}\left(\frac{R_e}{a}\right)^5 n_p \qquad (11)$$

where $Q_p$ is tidal quality factor, and $m_m$ is the mass of the moonlets. Since here we are studying the tidal evolution of the moonlet, we have to make a few assumptions on its mass.

This formula gives a semi-major axis changing rate of about 4m per century, far from enough to explain a drift from the equilibrium point if the satellite has been captured or formed there. It can also be seen from here that the effect of the Petit-Prince's mass on the mutual orbit is marginal. Concerning the Yarkovsky effect, we can compare the satellites of 45 Eugenia to those of Mars. This effect depends mainly on the distance between the Sun and the small objects considered. In the case of Phobos and Deimos, the diurnal Yarkovsky effect is negligible on the evolution of the satellites, a few cm of semi-major axis change on one million years (Tajeddine et al. 2011). 45 Eugenia being even farther from the Sun, we can safely neglect the Yarkovsky effect on the evolution of 45 Eugenia's satellites. The mechanism preventing the satellites from reaching the equilibrium points, or drifting them from it, is still to be determined, but their proximity to these points is a clear indication that these points are still important in the dynamics of the satellites and are good approximation of their position.

## 6. Conclusions

In this study, the potential locations of asteroidal moonlets with quasi-circular mutual orbit are investigated. It is assumed that the moonlet is a particle with infinitesimal mass. Both the solar gravity perturbation and the primary's 2nd degree-and-order gravity field are modeled. By analyzing the frozen solutions of the averaged equations of motion, we found that the orbits of several actual moonlets lie close to the classical Laplace equilibria, which reach global minimum values of the averaged Hamiltonian. The normal of the mean orbital plane of the moonlet, the primary's spin axis, and the normal of the primary's orbital plane around the Sun are found to be approximately coplanar, which is generally consistent with the previous studies (Boué & Laskar 2009; Fahnestock & Scheeres 2008). Even though no clear mechanism can explain the small difference between the satellites current position and the equilibrium points, they are good enough approximation for the satellites position. The position of these points do not depend on any a priori hypothesis on the moonlet's shape or mass apart from the fact that its mass is negligible with respect to the primary.

To determine those equilibrium positions, we need to know the orientation of the primary's spin pole, the primary's mass, the $J_2$ coefficient, and the moonlet's orbital size. The orientation of the primary's spin pole can be estimated from light-curve inversion in the case of 45 Eugenia for example (Taylor et al. 1988). Prior to the discovery of a satellite or a probe's fly-by, there is no possibility to determine precisely the mass of the primary. Yet, from the spectra, we can make assumptions on the primary's density and hence its mass. Its lightcurve can then provide its shape (Carry et al. 2012) and its polar oblateness (Turcotte & Schubert 2002). Most high ratio systems being compact, we can assume that the satellite would be at most at a distance of a few percent of the primary's Hill radius. A supposed semi-major axis in this range would then be a good first approximation. A systematic investigation around these equilibrium points may then lead us to discover these satellites.

**Acknowledgments**

This work was supported by National Basic Research Program of China (973 Program)